# Emerging Media Use and Acceptance of Digital Immortality: A Cluster Analysis among Chinese Young Adults

**Yi Mou**
School of Media and Communication
Shanghai Jiao Tong University
Shanghai, China
ORCID: 0000-0002-6836-5933

**Jianfeng Lan***
School of Media and Communication
Shanghai Jiao Tong University
Shanghai, China
jeff_lan@sjtu.edu.cn
ORCID: 0009-0000-1241-3361

**Jingyao Lu**
School of Media and Communication
Shanghai Jiao Tong University
Shanghai, China
ORCID: 0009-0003-2131-9504

** Corresponding author*

**Abstract:** Digital immortality is increasingly discussed as a technological possibility, yet empirical evidence about potential users' evaluations remains limited. We surveyed 462 Chinese young adults, combining cluster analysis of four emerging-media use frequencies with valence coding of open-ended responses to physical death, physical immortality, digital immortality, and digital death. Three profiles emerged: broad emerging-media, gaming-focused, and low-use users. Broad emerging-media users reported the highest adjusted acceptance and scored higher on several personality and worldview measures, while fear of death did not differ across profiles. Physical immortality elicited the most negative responses; digital immortality produced more mixed, less negative appraisals. More favorable digital-immortality appraisals predicted higher acceptance after adjustment for media-use profile and demographics, although scenario valence did not differ across profiles. The findings distinguish general receptivity associated with emerging-media repertoires from emotional responses to specific imagined futures, showing that evaluations depend on the form of continuity envisioned.

In the science fiction series Upload (Daniels, 2020–2025), a young programmer on the brink of death has his consciousness transferred into a digital cloud, allowing his psyche to transcend physical confines and his relationships to continue. Set in a near-future 2033, this narrative portrays a world where death is not an end, but a transition to a digital existence. While once confined to fiction, rapid advancements in artificial intelligence, big data, and brain-computer interfaces (BCI) have prompted renewed discussion of digital immortality's technological and ethical possibilities (Cebo, 2021; Galvão et al., 2021).

Digital immortality, broadly defined as the preservation of a person's identity and consciousness in a digital format after death (Bell & Gray, 2001), has sparked intense discussions. These conversations often center on the profound cultural, legal, and ethical dilemmas that arise when technology intersects with an experience as deeply ingrained in human culture as death (Cebo, 2021). However, the existing literature is predominantly theoretical, focusing on philosophical debates and the feasibility of technological implementation. There remains a significant gap in the empirical exploration of public attitudes towards this emerging concept, particularly from the perspective of potential users.

To address this gap, this study investigates digital immortality among young Chinese adults at two complementary levels. An exploratory cluster analysis identifies groups based on their use of related emerging media technologies and examines how those repertoires are associated with general acceptance, psychological characteristics, and worldviews. The study then analyzes participants' open-ended appraisals of four scenarios spanning physical death, physical immortality, digital immortality, and digital death. This second layer distinguishes a general evaluation of the technology from the concrete scenes and feelings it evokes, providing a fuller account of how potential users approach technologically mediated continuity.

## Literature Review

### Concept and Development of Digital Immortality

As an integral part of the natural life cycle, death has historically remained a social taboo. It was not until the 1950s that the fields of sociology and psychology began to offer preliminary insights into modern humanity's renewed engagement with the concept of death (Ariès, 1974). With the rapid advancement in AI, big data and related technology, our digital data have been utilized to emulate life before and after death (Galvão et al., 2017). Rahnama et al. (2021), for instance, explored the development of digital replicas through the 'Augmented Eternity and Swappable Identities' project.

Bell and Gray (2001) defined digital immortality as the continued active presence of a person's digital identity after death. Savin-Baden et al. (2017) further distinguished one-way and two-way forms. One-way digital immortality preserves materials through which others remember the deceased, including intentional memorials and digital traces that remain without prior planning. Two-way digital immortality adds interaction: a data-driven persona, chatbot, or avatar can continue communicating after the person's death. These forms primarily concern immortality for others. The present study examines the more ambitious consciousness-transfer form of immortality for oneself, following Cheng's (2023) description of storing or transferring personality or consciousness into a computer, robot, or cyberspace.

Advocates maintain the belief that immortality can be attained through the creation of non-biological "copies of the brain" (McGee & Maguire, 2007). However, it is crucial to acknowledge that the discourse on digital immortality is not uniformly positive. A growing body of critical scholarship highlights significant ethical, social, and economic concerns. An emerging industry is focused on the monetization and commercialization of the digital remains of the dead, raising profound questions about data ownership, privacy, and the potential for exploitation (Öhman & Floridi, 2018). There are legitimate fears that the deceased's vast and private data could be accessed and misused, or that their AI-generated avatar might inadvertently reveal secrets they intended to keep private (Harbinja, 2017; Kohl, 2022). These critical perspectives underscore the need for a cautious and nuanced examination of public attitudes toward this technology.

Existing research on digital immortality primarily focuses on the feasibility of technology implementation, cognitive issues (Savin-Baden & Burden, 2019), and ethical concerns (Galvão et al., 2021). Little attention has been paid to the public's attitudes towards it from the perspective of users as stakeholders.

### Technology Cluster and Adoption of Innovation for Digital Immortality

The tradition of technology adoption is deeply ingrained across a multitude of disciplines (Rogers, 2003). The most prominent key factors of an innovation are probably the five characteristics proposed by Rogers (2003), including relative advantage, compatibility, complexity, trialability, and observability.

Despite the considerable explanatory power of those five factors, certain aspects of innovations remain unaccounted for, such as the interdependence between an innovation and its technical predecessors, as no innovation exists in isolation. Rogers (2003) has coined the term "technology clusters" to delineate the boundaries around an innovation. By definition, a technology cluster "consists of one or more distinguishable elements of technology that are perceived as being closely interrelated" (Rogers, 2003, p. 14). In the trend of media technology convergence, scholars found that functionally similar technology clusters are significant predictors of adoption for new media like cable television and online radio (Lin, 2009). Notably, technology clusters are also indicative of an individual's level of innovativeness (Campagna & Bhada, 2024; Hunt et al., 2014).

Given the complexity of today's science and technology, it is evident that technologies are intricately interwoven (Ling, 2023), and the pursuit of digital immortality is no exception. The idea of digital immortality has evolved in tandem with the advancement of pertinent technologies. While memories may fade with time, digital data persist indefinitely. Concurrently, technologies like brain-computer interfaces (BCI) aim to foster direct communication between computers and the human brain, illustrating ongoing BCI development relevant to the broader technological context (Neuralink, 2023).

The four media were selected because they provide publicly accessible components of the broader technological ecology surrounding digital immortality. Virtual humans and AI chatbots instantiate responsive digital agents, whereas XR systems and video games provide interactive environments in which embodied or persistent digital identities can be experienced. None is equivalent to consciousness transfer. Their relevance lies in offering partial, currently available encounters with artificial agents, virtual embodiment, and digitally sustained identity, making repertoire-level use a plausible correlate of receptivity to a still-hypothetical technology.

Nevertheless, it is essential to acknowledge that the intricate scientific and technological underpinnings of digital immortality are not universally comprehended, and access to certain technological clusters remains limited for the average individual. For this study, we consider four relevant media technologies as the accessible technology cluster for digital immortality for the public: virtual humans, AI chatbots, XR technology, and video games. We expect the use of those technology clusters to be associated with the acceptance of digital immortality.

**Imagining Physical and Digital Forms of Death and Immortality**

Acceptance measures summarize whether a technology is evaluated favorably, but they cannot show whether people respond similarly to different forms of continued existence and disappearance. Physical death, physical immortality, digital immortality, and digital death locate continuity in different relations among the body, consciousness, and digital traces. Treating these states as interchangeable would obscure the possibility that a participant welcomes digital continuity while rejecting bodily permanence, or views the erasure of digital traces differently from bodily death. These distinctions are especially relevant to digital immortality, whose imagined benefits coexist with concerns about identity, privacy, and the posthumous use of personal data (Galvão et al., 2021; Öhman & Floridi, 2018).

The open-ended format also separates two judgments that a closed acceptance scale tends to combine: the valence of the scene a participant imagines and the feeling the participant reports in response to that scene. A digital future may appear socially or technologically intriguing while remaining personally unsettling. Examining both judgments across the four scenarios therefore clarifies whether receptivity to digital immortality reflects a generalized wish to defeat death or a more selective evaluation of how continuity would be achieved.

**Predictors of Digital Immortality Acceptance**

As cluster analysis is primarily employed for exploratory and hypothesis-generating endeavors (Shensa et al., 2018), no predefined hypotheses were established for this study. Concretely, we scrutinize three distinct categories of variables: individuals' views of death, levels of religiosity and spirituality, and their personality traits.

### *Fear of Death*

Confronting death, the anxiety would urge people to employ some defense mechanisms. In addition to approaches of attempting to achieve longevity, a sense of symbolic immortality is a prevalent strategy to mitigate the terror of death (Florian & Mikulincer, 1998). Coined by Lifton (1973), symbolic immortality manifests through five modes of experience.

Notably, this form of immortality departs from the symbolic and takes on a tangible dimension. Consequently, the fundamental question arises: Does the fear of mortality necessarily result in an embrace of digital immortality?

### *Religiosity and Spirituality*

Religions have long been described as “complicated systems of preparing for death” (Jung, 1969, p. 408). The relationship between religiosity and an individual’s attitude toward death is complex, with multiple competing theories. The buffering theory (Freud, 2012) and terror-management theory (Greenberg et al., 1986) both posit that religion alleviates the fear of death by espousing the existence of an afterlife, a claim supported by some empirical evidence (Norenzayan & Hansen, 2006; Vail et al., 2010). Conversely, other theories propose a curvilinear relationship, where the moderately religious experience the most death anxiety (Nelson & Cantrell, 1980), or even a positive correlation when individuals believe in a demanding deity (Ellis & Wahab, 2013).

Given this complexity, the role of religiosity in accepting digital immortality would be further uncertain, especially in Chinese society. Despite the contemporary claim of atheism in China, the nation has been significantly influenced by religions such as Buddhism and Daoism. Religious practices nevertheless remain part of China’s cultural landscape. Would those religious adults differ from their nonreligious counterparts in terms of accepting digital immortality? This question warrants further exploration.

In contrast to religiosity, spirituality reflects an internal and individual experience that operates without defined collective boundaries (Hafeez & Rafique, 2013). Unlike organized religious structures, spirituality allows individuals to “stand outside of their immediate sense of time and place to view life from a larger, more objective perspective” (Piedmont, 1999, p. 988). Yet, the principle of digital immortality appears to juxtapose with the essence of spirituality. Hence, it is imperative to investigate the potential relationship between spirituality and the acceptance of digital immortality.

### *Personality Traits*

Personality is widely regarded as one of the fundamental traits of humans (McAdams & Pals, 2006). The most widely used model is the Big Five model which categorizes personality traits into five dimensions, namely openness, conscientiousness, extraversion, agreeableness, and neuroticism (Costa & McCrae, 1992).

Furthermore, personality traits have been linked to technology acceptance more broadly. Prior investigations have consistently revealed a significant linkage between an individual’s innovativeness and the core personality trait of openness to experience (Marcati et al., 2008). Given the novelty of digital immortality technology, personality traits may therefore be relevant to individuals’ willingness to consider it. Thus, this study incorporates personality traits as one dimension of our investigation.

Existing theories on technology adoption, especially the concept of technology clusters, provide a robust framework for investigating this nascent area. However, it is unclear how the use of related media technologies, combined with fundamental individual differences, relates to acceptance of such a profound innovation.

Therefore, to address this gap, this study poses the following research questions:

RQ1: Based on their use of emerging media (virtual humans, AI chatbots, XR, and video games), what distinct user clusters can be identified among young Chinese adults?

RQ2: Do these user clusters differ significantly in their level of acceptance toward digital immortality?

RQ3: What are the demographic, psychological (personality traits, fear of death), and worldview (religiosity, spirituality) characteristics of each cluster?

RQ4: How do participants appraise the imagined scenes and their own feelings across physical death, physical immortality, digital immortality, and digital death, and how are appraisals of digital immortality associated with general acceptance?

# Method

## Study Design

This study employed a cross-sectional online survey design with closed-ended measures and open-ended scenario responses. An exploratory cluster analysis first classified participants according to their use frequencies of four emerging media technologies. Multinomial regression, ANOVAs, and an adjusted acceptance model were then used to characterize the clusters. Finally, valence-coded responses to four death-and-immortality scenarios were compared within participants and related to both cluster membership and general acceptance of digital immortality.

## Sample

An online survey was conducted on a purposive sample. Targets were recruited through a commercial online survey service Tencent Questionnaire (https://cloud.tencent.com/product/survey), which was contracted to send a recruitment announcement to its national sampling pool of more than 1 million adults. We specifically set the selection criteria as adults aged 18–44 years. Data collection occurred from May 23 to June 5, 2023.

We focused on younger adults because age is associated with attitudes toward mortality (Cicirelli, 2006), and younger cohorts have extensive opportunities to encounter virtual agents, interactive worlds, and persistent digital identities. They are therefore a theoretically relevant population for examining how current media repertoires relate to evaluations of a prospective technology. This sampling decision does not imply that all younger adults are technologically experienced or that they will necessarily become early adopters.

An online consent form was provided and signed before participation. The consent form emphasized anonymity, and the dataset used for analysis contained no direct personal identifiers. The study was approved by the institutional review board of the authors' university (No. REMOVED FOR PEER REVIEW).

A total of 1,027 potential participants were invited to participate in this survey, and 462 provided valid responses, yielding a response rate of 45%, higher than that of regular online surveys of 33–36% (Daikeler et al., 2022; Wu et al., 2022). Among all the 462 participants, 230 were females (49.8%) and the rest were males (50.2%). The average age was 23.69 years (SD = 6.65; three participants did not report their age, so age-related analyses are based on 459 participants). The monthly income level of 38.1% of participants (n = 176) was between CNY 5,001 to CNY 10,000, while the incomes of 30.3%, 15.8%, 7.1%, and 8.7% of the participants were below CNY 5,000, between CNY 10,001 and CNY 15,000, between CNY 15,001 and CNY 20,000, and more than CNY 20,000 (1 CNY equals about .14 USD), respectively. As for education level, 72.6% (n = 335) of participants had an education level of bachelor's degree or higher. In terms of religion, 65.6% (n = 303) claimed to be atheist, and the remaining 159 (34.4%) claimed to have some sort of religious beliefs.

**Measures**

Most of the original scales were from English literature and back-translated by two bilingual authors to ensure their accuracy, unless it was otherwise stated.

***Emerging Media Use Frequencies (Clustering Variables)***

Four types of emerging media were selected based on the criteria of 1) technology cluster for digital immortality, and 2) accessible to the public. They include 1) virtual humans, such as VTubers, virtual idols, and virtual anchors; 2) AI chatbots, such as ChatGPT chatbot and social bots on X (Twitter); 3) virtual reality (VR), augmented reality (AR), and mixed reality (MR); and 4) video games. Prior trace-data research on problematic smartphone use compared duration, frequency, and time distortion (Marciano & Camerini, 2022); accordingly, frequency was used as the clustering variable in this study. Participants first answered a screen question for each medium (1 = yes, 0 = no). Conditional frequency items were anchored at 1 = never and 7 = always. Screen nonselections and the resulting structurally missing conditional responses were coded 0 in the analyzed data, whereas observed conditional responses retained their 1–7 values. Thus, the analyzed frequency variables ranged from 0 to 7, and 0 should not be interpreted solely as never use.

***Acceptance of Digital Immortality (Dependent Variable)***

The participants' acceptance of digital immortality was measured with van der Laan et al.'s (1997) technology-acceptance scale. The instrument was originally developed for advanced transport telematics, but its semantic-differential format captures general evaluative responses to an unfamiliar technology. It was used as an exploratory measure because no validated scale specific to consciousness-transfer digital immortality was available when the survey was designed. Before responding, participants read a standardized definition describing digital immortality ("digital life") as the transfer and preservation of consciousness and memories in a computer system so that a person can continue to exist in a digital world, illustrated with a scene from the film *The Wandering Earth II*. The complete English translation and original Chinese wording are provided in the Supplemental Materials. The nine items used a 7-point response format with the unfavorable pole on the left and favorable pole on the right (e.g., "useless-useful" and "sleep-inducing-exciting"). Items were coded from 1 to 7; overall acceptance was their arithmetic mean. Cronbach's alpha was .97.

***Scenario-Based Open-Ended Appraisals***

Participants responded to four short scenarios presented in the following order: physical death, defined as bodily death; physical immortality, described as the pursuit of bodily permanence; digital immortality, in which the body dies but consciousness is converted into digital signals and stored permanently online; and digital death, in which bodily death is accompanied by the erasure of social-media accounts, data, and all other digital traces. The elicitation of valenced death imagery was informed by prior work on images of death and death attitudes (López-Pérez et al., 2020), while the digital-death scenario drew on scholarship distinguishing bodily death from the persistence or deletion of digital remains (Biçer & Yıldırım, 2022; Pitsillides et al., 2013). For each scenario, participants briefly described the scene they imagined and how they felt about it. Full translated prompts and original Chinese wording appear in the Supplemental Materials.

Each description was assigned two valence scores, one for the imagined scene and one for the participant's reported feeling, using a five-category scheme (−2 = strongly negative, −1 = negative, 0 = neutral or uncertain, 1 = positive, 2 = strongly positive). The retained analytic scores corresponded to the primary-coder ratings wherever paired coding records were available. A second coder independently rated the paired available responses for reliability assessment (n = 391–405 across the eight variables). Unweighted Cohen's κ ranged from .75 to .88. Depending on the scenario and judgment, 393–412 responses contained usable coded material. These variables represent emotional valence rather than thematic categories.

***Fear of Death***

Ho et al.'s (2010) 7-item fear subscale was adopted from their Chinese version of the Death Attitude scale to measure the fear of death. Participants were asked to indicate their degree of agreement from 1 (strongly disagree) to 7 (strongly agree). An example is "I have strong fear to death." The reliability coefficient Cronbach's α was .95.

***Religiosity***

The participants' degree of religiosity was gauged with Lo et al.'s (2010) Chinese Worldview Scale, which assesses Buddhist and Daoist orientations. Participants reported agreement with 12 items from 1 (strongly disagree) to 7 (strongly agree). Example items included "To achieve true happiness, one must accept the unavoidable facts of old age, sickness, and death" for Buddhism and "Natural objects and processes are full of wisdom; they can provide very good models for our lives and conduct" for Daoism. Cronbach's alpha was .86 for the six-item Buddhism subscale and .88 for the six-item Daoism subscale.

***Personality Traits***

Since the Big Five personality inventory has been translated into multiple languages and applied successfully cross-culturally, the Chinese version of the Big Five Personality Inventory-15 was adopted from Zhang et al. (2019). The items were adapted to a 7-point Likert response format, and participants reported on 15 items from 1 (strongly disagree) to 7 (strongly agree), three items for each factor. The reliability coefficient Cronbach's α was .89, .76, .86, .91, and .92 for the five factors of neuroticism, conscientiousness, agreeableness, openness, and extraversion, respectively.

***Spirituality***

The participants' degree of spirituality was measured by Piedmont's (1999) universality scale. The participants were asked to indicate how much they agreed with nine statements, such as "There is a higher plane of consciousness or spirituality that binds all people." The reliability coefficient Cronbach's α for the nine items was .90.

***Demographics (Covariates)***

Demographics, including age, sex, education level, and monthly family income level, were gauged as covariates using conventional measures. Their religious beliefs were also asked, and the data were further coded into atheist and non-atheist.

**Data Analysis**

Analyses proceeded in four stages. First, a k-means cluster analysis was performed on the z-standardized use frequencies of the four emerging media. The number of clusters was determined by the elbow criterion, parsimony, and substantive interpretability, with silhouette coefficients inspected as a descriptive check (Figure 1); a sensitivity analysis comparing k = 2–6 is reported in the Supplemental Materials (Table S1). K-means++ initialization with 100 random starts and a fixed random seed was used.

Second, a multinomial logistic regression modeled the associations of demographics and digital-immortality acceptance with nominal cluster membership, using broad emerging-media users as the reference. Third, one-way ANOVAs with Tukey HSD tests and partial eta-squared examined cluster differences in personality, religiosity, spirituality, and fear of death. An ANCOVA re-examined acceptance differences after adjustment for sex, age, education, and income. Fourth, because the scenario codes were ordinal and repeated within participants, Friedman tests compared the four scenarios separately for imagined scenes and reported feelings; paired Wilcoxon signed-rank tests used Holm-adjusted p values. Kruskal–Wallis tests examined cluster differences in the eight codes, with Holm adjustment across tests. Spearman correlations related the two digital-immortality codes to acceptance. Separate exploratory OLS models with HC3 robust standard errors estimated each association after adjustment for cluster, sex, age, education, and income.

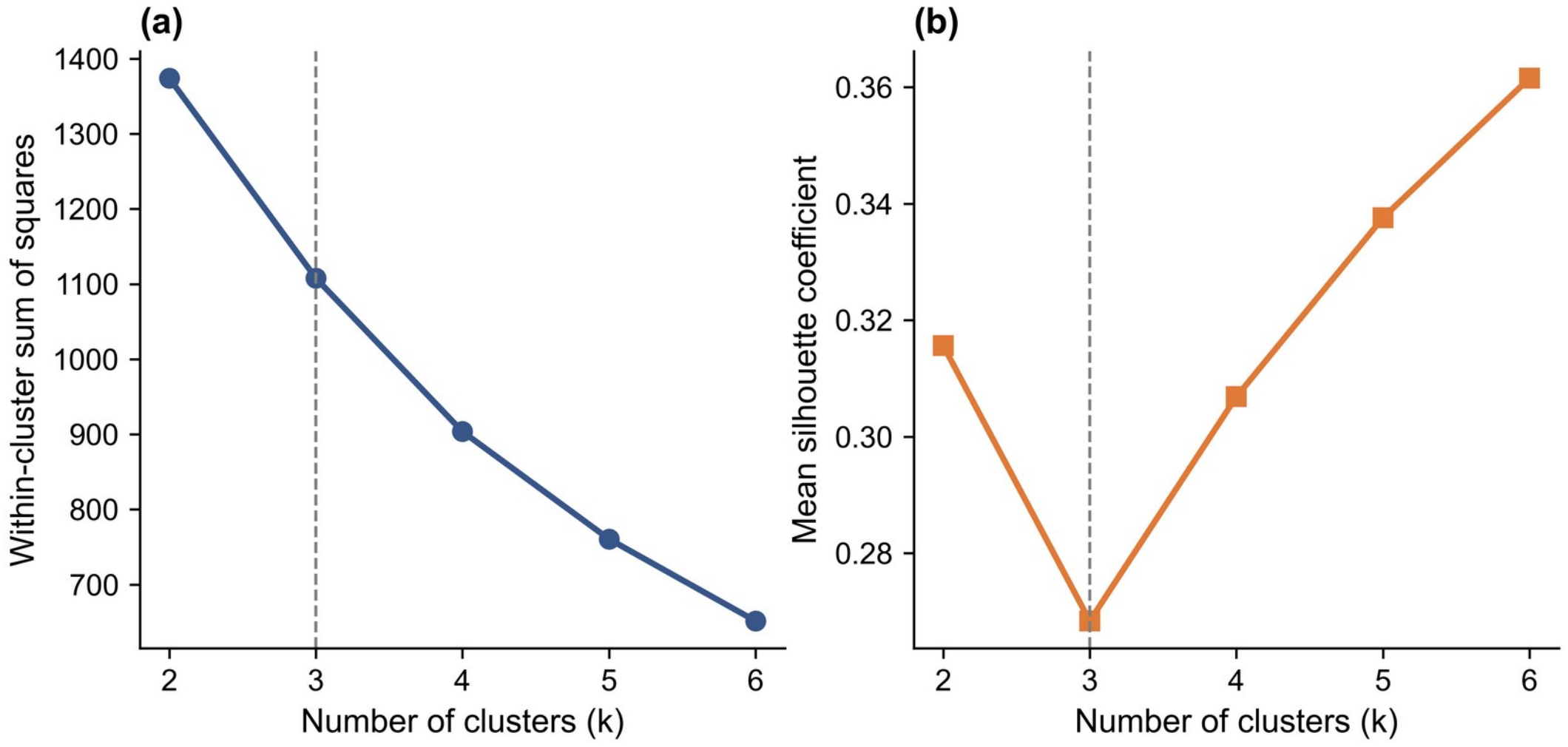

**Figure 1. Cluster-number determination**

*Note.* (a) elbow plot of within-cluster sum of squares and (b) mean silhouette coefficient for k = 2–6 clusters. The dashed line marks the selected three-cluster solution.

## Results

### Participants and Cluster Solution

All the 462 participants provided complete data on the four clustering variables. There was no evidence of multicollinearity among the clustering variables, with pairwise correlations ranging from −.04 to .28. The elbow plot (Figure 1a) showed diminishing returns beyond three clusters, and silhouette coefficients were modest across all solutions (range .27–.36 for k = 2–6; Figure 1b), indicating that no solution produced strongly separated clusters; the silhouette was therefore treated as descriptive rather than decisive. We retained the three-cluster solution because it was the most parsimonious solution that distinguished all substantive user profiles, and the acceptance ordering was stable across solutions with two to six clusters: the broadest-use cluster had the highest acceptance and a low-use cluster had the lowest (Table S1 in the Supplemental Materials). Hence, the 462 participants were classified into three clusters (see the composition of clusters by emerging media use frequency in Table 1 and Figure 2).

**Table 1. Composition of Clusters by Emerging Media Use Frequency**

| Cluster Label | N (%) | VR/AR/MR | Video Game | AI Chatbot | Virtual Human |
|---|---|---|---|---|---|
| Cluster 1: Broad emerging-media users | 89 (19.3%) | 2.73 (2.47)[a] | 3.93 (2.46)[a] | 2.80 (2.24)[a] | 3.93 (1.54)[a] |
| Cluster 2: Low-use users | 168 (36.4%) | 1.01 (1.74)[b] | 0.92 (1.26)[b] | 0.56 (1.25)[b] | 0.24 (0.74)[b] |
| Cluster 3: Gaming-focused users | 205 (44.4%) | 0.86 (1.57)[b] | 5.29 (1.29)[c] | 1.50 (1.98)[c] | 0.32 (0.77)[b] |
| *F(2, 459)* | | 34.83*** | 362.23*** | 44.91*** | 521.96*** |

*Note.* Values are M (SD). Use frequency ranged from 0 to 7. Conditional frequency responses used 1 = never and 7 = always; screen nonselections and missing conditional frequency responses were coded 0, so 0 should not be interpreted solely as never use. Superscripts a, b, c denote significant pairwise differences (Tukey HSD, $p < .05$). *** $p < .001$.

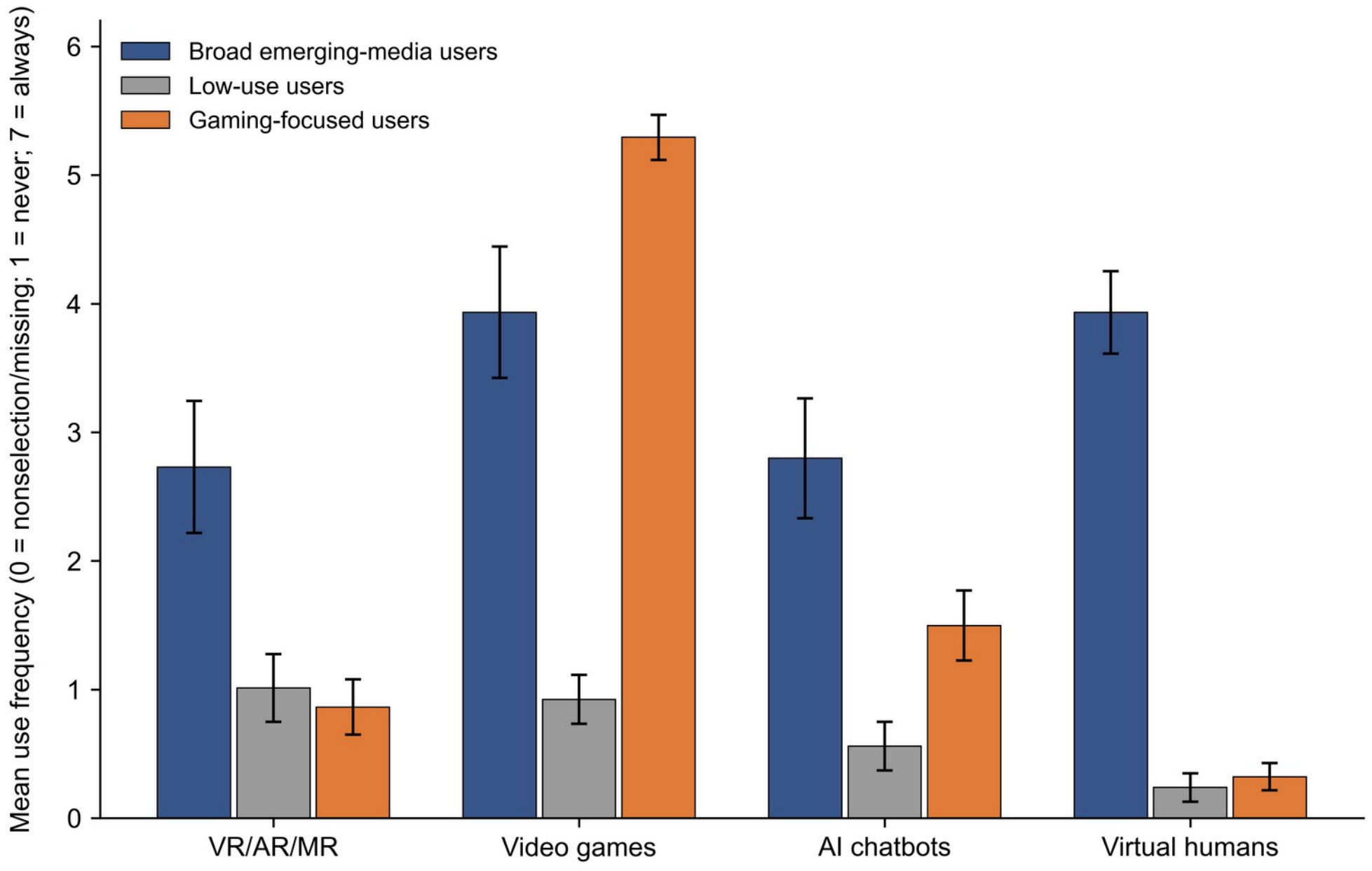

**Figure 2. Emerging-media use frequency by cluster**

*Note.* Error bars represent 95% confidence intervals; coding is 0 for screen nonselection or missing conditional response and 1–7 for conditional frequency responses (1 = never, 7 = always).

## Description of Clusters

The first cluster was labeled “broad emerging-media users” (see Table 2). This cluster (n = 89, 19.3%) had the highest use frequency of all emerging media technologies except video games. Their use frequencies of VR/AR/MR (M = 2.73, SD = 2.47), AI chatbots (M = 2.80, SD = 2.24), and virtual humans (M = 3.93, SD = 1.54) were significantly higher than those of the other two clusters. The percentage of atheists was the lowest in this cluster (53.9%), compared to 64.9% and 71.2% in the other two clusters.

The second cluster was labeled “low-use users.” This cluster (n = 168, 36.4%) exhibited the lowest use frequency of all four emerging media technologies. Notably, the percentage of males was the lowest (37.5%), relative to 52.8% in the broad emerging-media users cluster and 59.5% in the third cluster.

The third cluster was labeled "gaming-focused users" (n = 205, 44.4%) since they played video games the most frequently (M = 5.29, SD = 1.29), while maintaining a low-to-moderate level in using the other three media. The omnibus test for age was significant but small, $F(2, 456) = 3.15$, $p = .044$, and no pairwise comparison reached significance; education and income did not differ significantly across clusters (see Table 2).

**Table 2. Demographic Composition of Whole Sample and Each Cluster**

| Demographics | Overall | *Broad media users* | *Low-use users* | *Gaming-focused users* | Test statistic |
|---|---|---|---|---|---|
| Age | 23.69 (6.65) | 22.89 (5.31) | 24.70 (7.86) | 23.21 (5.97) | $F(2, 456) = 3.15$* |
| Sex (male %) | 50.2 | 52.8[a] | 37.5[b] | 59.5[a] | $\chi^2(2) = 18.19$*** |
| Education | 3.74 (0.83) | 3.65 (0.85) | 3.83 (0.86) | 3.70 (0.78) | $F(2, 459) = 1.80$ |
| Income | 2.26 (1.21) | 2.35 (1.16) | 2.25 (1.28) | 2.22 (1.18) | $F(2, 459) = 0.33$ |
| Atheist (%) | 65.6 | 53.9[a] | 64.9[ab] | 71.2[b] | $\chi^2(2) = 8.27$* |

*Note.* N = 462; age based on n = 459. Values are M (SD) or percentages. Broad media users = broad emerging-media users; Gaming-focused = gaming-focused users. Education and income were measured on 5-point ordinal scales. Superscripts a, b denote significant pairwise proportion differences ($p < .05$). * $p < .05$. *** $p < .001$.

### Cluster Differences in Acceptance of Digital Immortality

A one-way ANOVA revealed a significant difference in the acceptance of digital immortality across the three clusters, $F(2, 459) = 13.83$, $p < .001$, $\eta^2p = .057$ (Figure 3). Tukey HSD tests showed that the broad emerging-media users cluster (M = 5.14, SD = 1.07) reported significantly higher acceptance than both the low-use users cluster (M = 4.28, SD = 1.29) and the gaming-focused users cluster (M = 4.47, SD = 1.33), whereas the latter two clusters did not differ significantly from each other ($p = .328$). This pattern was robust to demographic adjustment: an ANCOVA controlling for sex, age, education, and income yielded adjusted means of 5.15, 4.28, and 4.46 for the broad emerging-media users, low-use users, and gaming-focused users clusters, respectively, and both contrasts against the broad emerging-media users cluster remained significant (low-use users: $b = -0.87$, 95% CI [−1.21, −0.54]; gaming-focused users: $b = -0.70$, 95% CI [−1.02, −0.38], ps < .001).

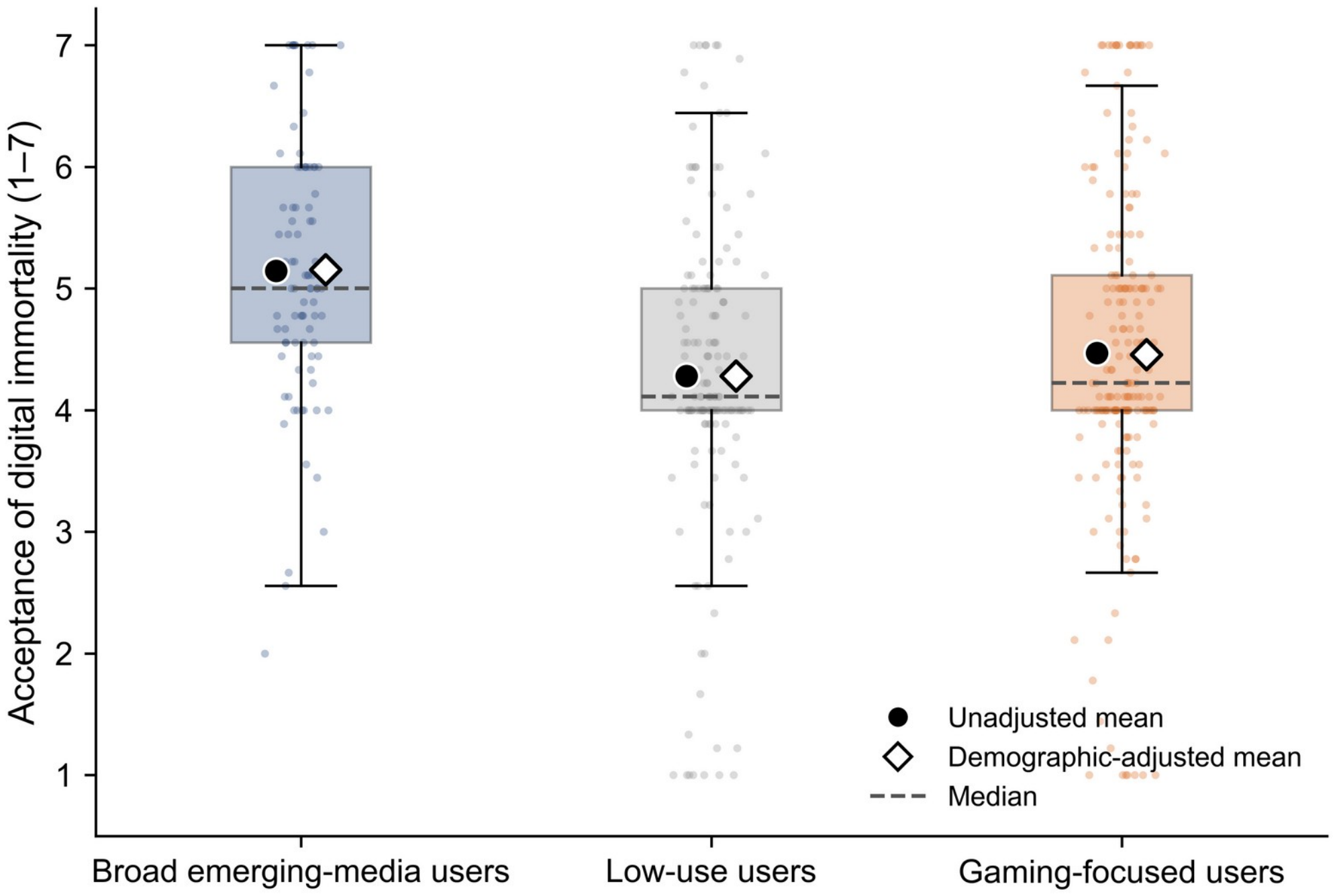

**Figure 3. Acceptance of digital immortality by cluster**

*Note.* Boxes show the interquartile range, dashed gray lines mark medians, black circles mark unadjusted means, and white diamonds mark means adjusted for sex, age, education, and income (ANCOVA).

**Multivariable Associations of Demographics and Acceptance with Cluster Membership**

A multinomial logistic regression was conducted to assess the multivariable associations between demographic characteristics, acceptance of digital immortality, and cluster membership, with the broad emerging-media users cluster as the reference category. The overall model was significant, $\chi^2(10) = 66.91$, $p < .001$ (pseudo $R^2 = .07$; N = 459). The results (see Table 3) confirmed the significant association between cluster membership and the acceptance of digital immortality. Relative to the broad emerging-media users cluster, each one-point increase in acceptance was associated with lower odds of belonging to the low-use users cluster: odds ratio = 0.55, 95% confidence interval = [0.44, 0.70]. Similarly, relative to the broad emerging-media users cluster, higher acceptance was associated with lower odds of belonging to the gaming-focused users cluster: odds ratio = 0.63, 95% confidence interval = [0.50, 0.78]. Demographically, being male was associated with lower odds of low-use membership (odds ratio = 0.47, 95% CI = [0.26, 0.84]), and age was associated with slightly higher odds of low-use membership (odds ratio = 1.07, 95% CI = [1.03, 1.12]).

**Table 3. Multinomial Logistic Regression Predicting Cluster Membership (Reference Category: Broad Emerging-Media Users)**

| Cluster | Sex (male) | Age | Education | Income | Acceptance |
|---|---|---|---|---|---|
| Broad emerging-media users | Reference | Reference | Reference | Reference | Reference |
| Low-use | **0.47 [.26, .84]*** | **1.07 [1.03, 1.12]**** | 1.37 [.96, 1.96] | 0.85 [.67, 1.08] | **0.55 [.44, .70]***** |
| users | | | | | |
| Gaming-focused users | 1.49 [.87, 2.58] | 1.01 [.96, 1.06] | 1.17 [.84, 1.63] | 0.91 [.73, 1.14] | **0.63 [.50, .78]**$^{***}$ |

*Note.* N = 459. Numbers are odds ratios with 95% confidence intervals in brackets. Overall model: $\chi^2(10) = 66.91$, $p < .001$; pseudo $R^2 = .07$. * $p < .05$. ** $p < .01$. *** $p < .001$.

## Characteristics of Clusters

The participants in different clusters exhibited different personality traits (see Table 4 and Figure 4). Broad emerging-media users were significantly more conscientious (M = 5.08, SD = 1.13) than the low-use users (M = 4.55, SD = 1.23) and the gaming-focused users (M = 4.58, SD = 1.32): $F(2, 459) = 6.20$, $p = .002$, $\eta^2p = .026$. Broad emerging-media users were also significantly more open (M = 4.68, SD = 1.39) and more extraverted (M = 4.21, SD = 1.58) than the other two clusters: $F(2, 459) = 8.55$, $p < .001$, $\eta^2p = .036$ for openness, and $F(2, 459) = 4.04$, $p = .018$, $\eta^2p = .017$ for extraversion. Neuroticism, $F(2, 459) = 1.06$, $p = .348$, and agreeableness, $F(2, 459) = 2.57$, $p = .078$, did not differ significantly across clusters.

In line with the pattern of atheism, broad emerging-media users exhibited significantly stronger beliefs in Buddhism (M = 5.52, SD = 0.97) and Daoism (M = 5.51, SD = 0.92) than the other two clusters: $F(2, 459) = 4.70$, $p = .010$, $\eta^2p = .020$ for Buddhism, and $F(2, 459) = 5.12$, $p = .006$, $\eta^2p = .022$ for Daoism. They also had significantly higher levels of spirituality (M = 5.50, SD = 0.89) than the other two clusters: $F(2, 459) = 8.98$, $p < .001$, $\eta^2p = .038$. However, the participants in those three clusters showed no significant difference in the level of fear of death: $F(2, 459) = 0.58$, $p = .558$.

### Table 4. Characteristics of Each Cluster

| **Characteristic** | **Overall** | *Broad media users* | *Low-use users* | *Gaming-focused users* | **F(2, 459)** | **$\eta^2$p** |
|---|---|---|---|---|---|---|
| Neuroticism | 4.20 (1.58) | 4.36 (1.63) | 4.07 (1.58) | 4.23 (1.56) | 1.06 | .005 |
| Conscientiousness | 4.66 (1.27) | 5.08 (1.13)[a] | 4.55 (1.23)[b] | 4.58 (1.32)[b] | 6.20** | .026 |
| Agreeableness | 4.72 (1.29) | 4.93 (1.23) | 4.56 (1.32) | 4.76 (1.27) | 2.57 | .011 |
| Openness | 4.13 (1.43) | 4.68 (1.39)[a] | 3.95 (1.38)[b] | 4.05 (1.44)[b] | 8.55*** | .036 |
| Extraversion | 3.78 (1.61) | 4.21 (1.58)[a] | 3.71 (1.65)[b] | 3.65 (1.57)[b] | 4.04* | .017 |
| Buddhism | 5.21 (1.09) | 5.52 (0.97)[a] | 5.10 (1.09)[b] | 5.16 (1.12)[b] | 4.70* | .020 |
| Daoism | 5.18 (1.11) | 5.51 (0.92)[a] | 5.06 (1.18)[b] | 5.14 (1.10)[ab] | 5.12** | .022 |
| Universality | 5.08 (1.07) | 5.50 (0.89)[a] | 4.95 (1.14)[b] | 5.00 (1.05)[b] | 8.98*** | .038 |
| Fear of death | 3.74 (1.66) | 3.84 (1.72) | 3.63 (1.66) | 3.78 (1.63) | 0.58 | .003 |

*Note.* Values are M (SD). Broad media users = broad emerging-media users; Gaming-focused = gaming-focused users. Superscripts a, b, c denote significant pairwise differences (Tukey HSD, * $p < .05$. ** $p < .01$. *** $p < .001$.

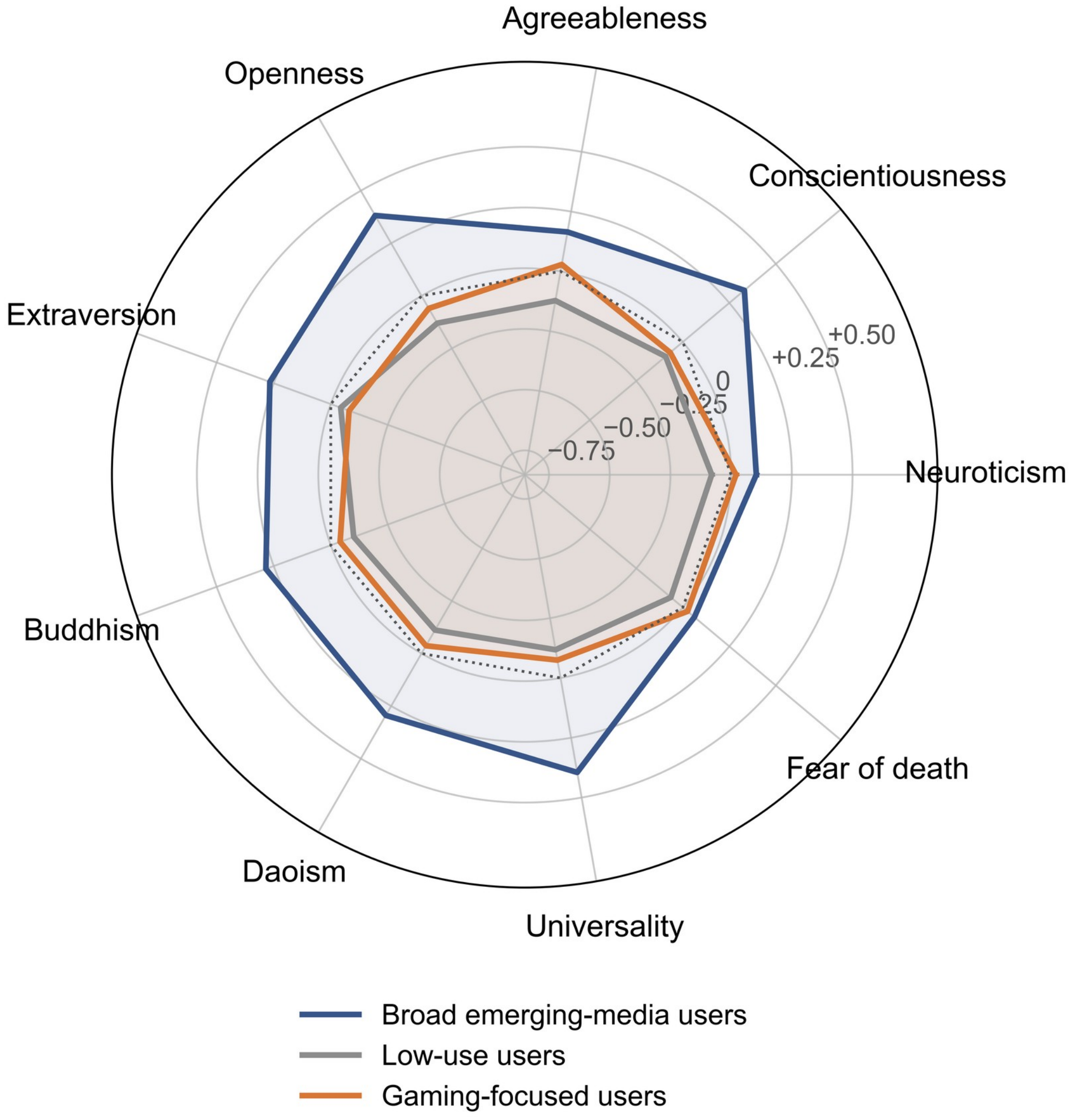

**Figure 4. Standardized psychological profile of the three clusters**

(z-scores of cluster means relative to the full sample)

### Emotional Appraisals across the Four Scenarios

Valid coding was available for 393–412 participants across the eight appraisal variables (Figure 5). Physical immortality elicited the most consistently negative responses: 64.0% of imagined-scene codes and 67.2% of feeling codes were negative or strongly negative. Digital immortality was more divided. Its imagined scenes were 30.8% negative, 38.9% neutral or uncertain, and 30.3% positive; the corresponding feelings were 38.5% negative, 33.7% neutral or uncertain, and 27.8% positive. Responses to physical death and digital death were concentrated in the neutral and negative categories, with relatively few positive codes.

The excerpts clarify what the valence categories represented without constituting a separate thematic analysis. Physical death could be framed neutrally as returning to nature (P013: “Dust returns to dust and earth to earth ... the common and equal ending of humankind”), whereas physical immortality could be rejected as the dissolution of embodied meaning (P011: “Humanity would enter a greater emptiness ... I resist that day”). Digital immortality generated sharply contrasting evaluations. One participant raised questions about whether an online consciousness would count as life and whether it could be stolen, exploited, or controlled (P019); another imagined preserved knowledge, closer intergenerational connections, and the continuation of human emotion (P050). Digital death was likewise interpreted as either painful erasure (P054) or release from the demand to remain recorded (P013). Additional translated excerpts and the original Chinese responses are provided in the Supplemental Materials.

The four scenarios differed for both imagined scenes, $\chi^2(3, N = 394) = 79.82$, $p < .001$, and reported feelings, $\chi^2(3, N = 393) = 93.09$, $p < .001$. Holm-adjusted pairwise tests showed that digital immortality was less negatively appraised than each of the other three scenarios for both judgments (adjusted ps ≤ .002). Physical immortality was more negative than physical death and digital death (adjusted ps < .001), while physical death and digital death did not differ. Thus, the contrast was not a simple preference for immortality over death; the form in which continuity or disappearance was imagined mattered.

None of the eight valence codes differed significantly across the media-use clusters (all raw ps ≥ .128; all Holm-adjusted ps = 1.00). Within the digital-immortality scenario, however, more positive imagined-scene and feeling codes were associated with higher general acceptance (Spearman's $\rho = .18$, $p < .001$, $n = 396$, and $\rho = .24$, $p < .001$, $n = 395$, respectively). In separate adjusted models, the imagined-scene code remained associated with acceptance, $b = 0.16$, 95% CI [0.05, 0.28], $p = .006$, $\Delta R^2 = .020$, as did the feeling code, $b = 0.21$, 95% CI [0.10, 0.33], $p < .001$, $\Delta R^2 = .038$. These associations indicate convergence between the scenario-specific coding and the acceptance scale while leaving substantial variation unexplained.

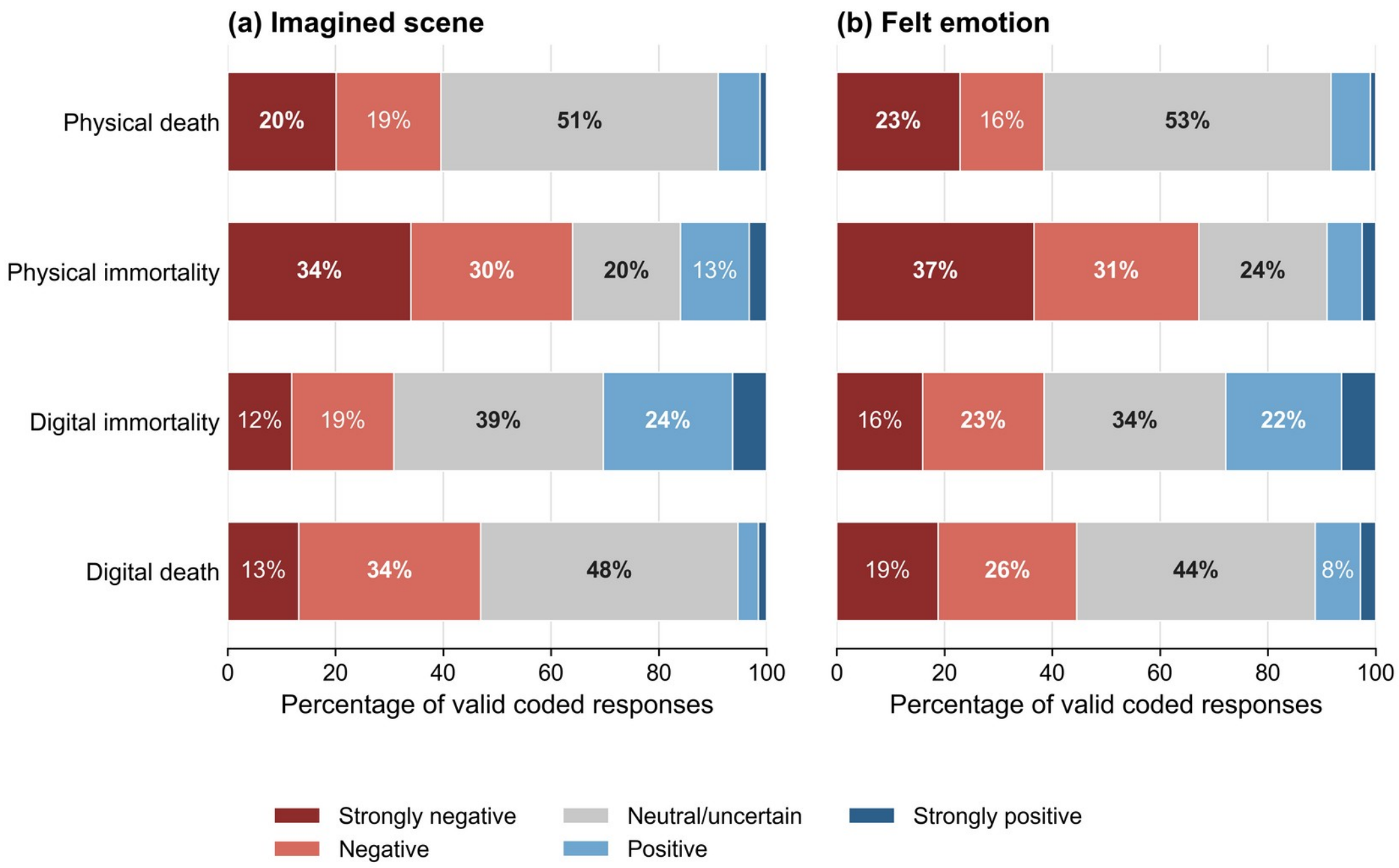

**Figure 5. Valence distributions for imagined scenes and reported feelings across four death-and-immortality scenarios**

*Note.* Percentages are based on valid coded responses for each scenario (n = 393–412). Codes ranged from −2 (strongly negative) to 2 (strongly positive).

## Discussion

Viewed through a media-psychology lens, the results show that young adults' emerging-media repertoires form distinct profiles associated with psychological characteristics and general receptivity to digital immortality. The three clusters differed in acceptance, and the pattern was robust to demographic adjustment. This result is consistent with the technology-cluster perspective (Rogers, 2003), according to which related technologies tend to be adopted together, and with evidence that use of related media clusters is associated with adoption of newer media (Lin, 2009).

Broad emerging-media users were more open, conscientious, and extraverted than the other clusters. Gaming-focused users and low-use users did not differ in acceptance, indicating that frequent gaming alone was not associated with the greater receptivity observed among participants who used a wider range of emerging media. Broad emerging-media users also reported stronger Buddhist and Daoist worldviews and greater spirituality. These cross-sectional associations do not establish whether media use shapes acceptance, whether receptive individuals seek out multiple technologies, or whether both reflect other dispositions.

The scenario results refine this cluster-level account. Participants reacted most negatively to physical immortality and treated digital immortality with markedly greater ambivalence. Digital immortality therefore did not function as a simple technological solution to fear of death. Many respondents could imagine it neutrally or positively, yet a substantial share still reported negative feelings. The significant associations between digital-immortality appraisals and general acceptance support the substantive relevance of these brief responses, while the absence of cluster differences in the valence codes shows that explicit technology acceptance and immediate scenario-based emotion are related without being equivalent. Media repertoires differentiated general receptivity; the concrete form of death or continuity organized the scenario appraisals more strongly.

While our study focused on the general concept of digital immortality, it is crucial to unpack the distinct ethical landscapes of “immortality for oneself” versus “immortality for others.” The latter, which involves creating digital replicas of the deceased, presents more immediate and tangible societal challenges. Digital-afterlife systems may also raise issues of inequality in access, digital literacy, and preparedness. Existing discussion of these disparities suggests a possible posthumous digital divide, in which differences in resources and skills could shape who can participate or be represented (Park et al., 2024). This risks turning a fundamental human experience into a stratified consumer product.

Finally, the psychological impact on the bereaved warrants careful consideration. While interacting with a deceased loved one’s avatar may offer short-term comfort, it could also foster an unhealthy attachment, preventing the necessary process of grieving and acceptance (Jiménez-Alonso & Brescó de Luna, 2023; Lindemann, 2022). This is particularly relevant given that the most receptive cluster also reported stronger Buddhist and Daoist worldviews and greater spirituality. This complex interplay between technology, grief, and faith highlights a critical area for future research as we navigate the profound implications of digital immortality.

## Limitations

The interpretation of these findings is subject to several limitations. The cross-sectional and exploratory design does not establish causal direction, and the exact cluster solution may vary with the sample and algorithmic choices despite the deterministic pipeline and sensitivity checks. The general acceptance scale was adapted from a broader technology-acceptance measure and may not capture all ethical or existential dimensions of consciousness-transfer technologies. The open-ended scenarios partly address this limitation, but their brief responses were reduced to coarse valence codes. They cannot support claims about themes, reasoning processes, or meanings that were not formally coded, and the usable sample varied modestly across appraisal variables. Finally, the purposive sample consisted of relatively young and highly educated Chinese adults. More representative, longitudinal, experimental, and cross-cultural research is needed to test whether the observed repertoire and scenario patterns generalize.

## Conclusion

This study provides an empirical account of digital-immortality acceptance that connects emerging-media repertoires with scenario-specific emotional appraisals. Broad emerging-media users were the most receptive on the general acceptance scale and differed from other participants on several personality and worldview measures. Yet the open-ended responses showed a pattern that cut across those clusters: physical immortality was evaluated most negatively, whereas digital immortality elicited a mixture of negative, neutral, and positive reactions. More favorable digital-immortality appraisals were associated with higher general acceptance, but the absence of cluster differences in the four-scenario codes indicates that media use and immediate emotional imagination capture distinct aspects of receptivity. Public responses to digital immortality therefore depend on both prior engagement with related technologies and the specific form of life, death, and continuity people are asked to imagine.

## Statements and Declarations

**Prior dissemination.** An earlier version of this manuscript was posted by the authors as a preprint at https://doi.org/10.48550/arXiv.2505.01355. The present manuscript has been substantially revised in its framing, terminology, methods reporting, robustness checks, figures, and references.

**Generative artificial intelligence disclosure.** A large language model (OpenAI GPT-5.6 Sol) was used solely for language editing. The authors reviewed and approved all AI-assisted revisions and take full responsibility for the manuscript content.

**Data availability.** The data and analysis code supporting the findings are available from the corresponding author upon reasonable request, subject to applicable ethical and privacy restrictions.

**Declaration of conflicting interest.** The authors declared no potential conflicts of interest with respect to the research, authorship, and/or publication of this article.

**Funding.** This study was supported by the Humanity and Social Science Planning Fund of Ministry of Education of China under Grant no. 25YJA860010.

**Ethics and consent.** The study was approved by the Institutional Review Board of Shanghai Jiao Tong University (approval no. H2022332I). All participants provided written informed consent through the online consent form before beginning the survey.

**Acknowledgments.** We thank Professor Jilong Wang for his advice and conceptual insights during the early development of this study.

# Supplemental Materials

## S1. Definition of Digital Immortality Presented to Participants

Immediately before completing the acceptance scale, all participants read the following standardized definition. The English translation is followed by the original Chinese wording used in the questionnaire.

*English translation:*

“Digital life refers to transferring information such as a person's consciousness and memories into a computer system in order to achieve 'immortality' in the digital world. This technology can transcend the body's natural processes of birth, aging, illness, and death, and the resulting digital entity has personality, thought, and autonomous consciousness like a human being. For example, in the film The Wandering Earth II, the scientist Tu Hengyu stores his daughter Yaya's consciousness in a quantum computer and ultimately succeeds in digitizing her life so that she can continue to exist in the digital world.”

*Original Chinese wording:*

“数字生命是指将人的意识和记忆等信息转移到计算机系统中，以实现在数字世界中的“永生”。这种技术可以脱离肉体生老病死的自然规律，具有和人一样的人格、思维和自主意识。例如在电影《流浪地球 2》中，科学家图恒宇将女儿丫丫的意识保存进量子计算机，最终成功使女儿的生命数字化并能够在数字世界中延续。”

**S2. Four Open-Ended Scenario Prompts**

***Physical death***

English translation: In the traditional sense, death means bodily death. What kind of scene do you think this would be? How would you feel about it? Please describe briefly.

Original Chinese: 传统意义上的“死亡”即“肉身死亡”。你认为这将会是一副怎样的景象？对此你有何种感受？请简单描述。

***Physical immortality***

English translation: Throughout history, many people have pursued bodily immortality. What kind of scene do you think bodily immortality would be? How would you feel about it? Please describe briefly.

Original Chinese: 古往今来，不少人在追求“肉体永生”。你认为“肉体永生”将会是一副怎样的景象？对此你有何种感受？请简单描述。

***Digital immortality***

English translation: Many people today discuss digital immortality: the body dies, but consciousness is converted into digital signals and stored permanently in online space. What kind of scene do you think this would be? How would you feel about it? Please describe briefly.

Original Chinese: 今天不少人在谈论“数字永生”，即肉身死亡，但意识转为数字信号，永久保存在网络空间中。你认为这将会是一副怎样的景象？对此你有何种感受？请简单描述。

***Digital death***

English translation: In addition to bodily death, suppose that all of a person's digital traces, including social-media accounts and other data, were erased so that no record of that person remained in the world. This would be considered digital death. What kind of scene do you think digital death would be? How would you feel about it? Please describe briefly.

Original Chinese: 除了肉体死亡外，如果一个人的所有数字痕迹，包括社交媒体账号，各种数据资料等等都将被抹去，世界上不再有过他/她的任何记录，那么这被视为“数字死亡”。你认为“数字死亡”将会是一副怎样的景象？对此你有何种感受？请简单描述。

## S3. Valence Coding and Intercoder Agreement

Each response received one code for the valence of the imagined scene and one for the participant's expressed feeling: -2 = strongly negative, -1 = negative, 0 = neutral or uncertain, 1 = positive, and 2 = strongly positive. The table reports usable analytic cases and agreement between the primary and second coders in the paired available records. Unweighted Cohen's kappa was used because the manuscript reports exact-category agreement.

**Table S1. Usable Cases and Intercoder Agreement for Scenario Valence Coding**

| Scenario | Judgment | Usable n | Paired n | Cohen’s kappa |
|---|---|---|---|---|
| Physical death | Imagined scene | 412 | 405 | .88 |
| | Feeling | 411 | 404 | .84 |
| Physical immortality | Imagined scene | 400 | 397 | .75 |
| | Feeling | 399 | 397 | .82 |
| Digital immortality | Imagined scene | 396 | 393 | .76 |
| | Feeling | 395 | 393 | .79 |
| Digital death | Imagined scene | 394 | 391 | .79 |
| | Feeling | 393 | 391 | .82 |

**S4. Illustrative Responses**

The excerpts below were selected to show variation within each scenario, not to establish qualitative themes or estimate their prevalence. Participant numbers are study-specific identifiers. English translations were prepared for reporting; the original Chinese wording is retained for auditability.

***Physical death, P013 (scene = 0; feeling = 0)***

English translation: “Dust returns to dust and earth to earth, decomposing until it returns to nature; the common and equal ending of humankind.”

Original Chinese: “尘归尘土归土，分解直至回归自然；人类的公平结局。”

***Physical death, P039 (scene = -1; feeling = -2)***

English translation: “Bodily death is the end of all personal meaning at every level; compared with physical pain, the fear one would face mentally makes me more uneasy.”

Original Chinese: “肉身死亡是个人一切意义在所有层面上的终结；相较于肉体痛苦，精神上将面临的恐慌更让我不安。”

***Physical immortality, P011 (scene = -2; feeling = -2)***

English translation: “With bodily immortality, the meaning of embodiment would dissolve and humanity would enter a greater emptiness; this may be where technology is heading, but I resist that day.”

Original Chinese: “肉体永生，具身性的意义消解，人类必将进入更大的虚无中；或许是技术所趋，但我抗拒这一天。”

***Physical immortality, P196 (scene = 1; feeling = 1)***

English translation: “Never aging would be bodily immortality, something visible and tangible; if one were not too old, it might be all right.”

Original Chinese: “长生不老就是肉体永生，看得见摸得着；如果不是太老还可以。”

***Digital immortality, P019 (scene = -1; feeling = -1)***

English translation: “If consciousness is online, does it count as life? Would it require human maintenance? Could it be stolen or exploited, stripped of rights, or controlled by others? It gives me a sense of technological crisis.”

Original Chinese: “意识在网络，算生命吗？需要人工维护吗？会被盗窃利用吗？会丧失人权为人控制吗？技术主义危机感。”

***Digital immortality, P050 (scene = 2; feeling = 2)***

English translation: “It could preserve rich thought and perspectives, connect knowledge across time, and retain outstanding people. Connections between people could become closer, linking ancestors and descendants; the world would become richer and human emotions could continue.”

Original Chinese: “展现出很丰富思维、眼界，将古往今来的见闻联通，也能保存优秀人才，同时人与人的联络会更加亲密，对自己的祖先到后代都有一定的了解；世界会变得更加丰富，人类的情感得到延续。”

***Digital death, P054 (scene = -1; feeling = -2)***

English translation: “Every trace of a person's existence would disappear, and no one could find what they had said or done; for the living, it would be extremely painful not to be able to find traces of someone they once loved.”

Original Chinese: “一个人存在的痕迹全部消失，没人可以查找到他曾说过什么话、做过什么事；我觉得对生者而言非常痛苦，无法寻觅曾经爱过的人存在的痕迹。”

***Digital death, P013 (scene = 1; feeling = 0)***

English translation: “A person would no longer be connected to the world and would be completely released. Over a long enough timeline, being forgotten is the final outcome of every person and civilization; once I can no longer feel it, whether I am remembered has nothing to do with me.”

Original Chinese: “人与世界再无关联，彻底解脱；拉长时间线，被遗忘是所有人、所有文明最终的结局，既然我已经感受不到，那是否能被记住与我无关。”

## S5. Sensitivity Analysis of Cluster Solutions

The retained three-cluster solution was compared with k = 2-6 solutions using the same z-standardized clustering variables, k-means++ initialization, 100 random starts, and fixed random seed.

**Table S2. Sensitivity Analysis of Cluster Solutions With k = 2-6 Clusters**

| k | Silhouette | ARI with k = 3 | Cluster sizes | Acceptance M (SD), broadest-use | Acceptance M (SD), lowest-use |
|---|---|---|---|---|---|
| 2 | .316 | .235 | 158, 304 | 4.90 (1.25) | 4.34 (1.30) |
| 3 | .268 | - | 89, 168, 205 | 5.14 (1.07) | 4.28 (1.29) |
| 4 | .307 | .512 | 67, 112, 130, 153 | 5.23 (1.14) | 4.25 (1.26) |
| 5 | .338 | .368 | 67, 70, 90, 95, 140 | 5.25 (1.13) | 4.19 (1.39) |
| 6 | .362 | .331 | 37, 50, 70, 79, 93, 133 | 5.45 (1.11) | 4.23 (1.37) |

Note. ARI = adjusted Rand index relative to the retained three-cluster solution. Across k = 2-6, the broadest-use cluster had the highest acceptance and a low-use cluster had the lowest. One-way ANOVAs of acceptance were significant for every solution, Fs(1-5, 456-460) = 5.94-20.36, ps < .001.